\documentclass[a4paper,12pt,feynmf]{article}
\usepackage{amssymb}
\usepackage{amsmath,amssymb,mathrsfs}
\usepackage{graphicx}
\usepackage{ifpdf}
\usepackage{xcolor}
\newcommand*{\rom}[1]{\expandafter\@slowromancap\romannumeral #1@}
\begin{document}
\def\ctr#1{\hfil $\,\,\,#1\,\,\,$ \hfil}
\def\tstrut{\vrule height 2.7ex depth 1.0ex width 0pt}
\def\mystrut{\vrule height 3.7ex depth 1.6ex width 0pt}
\def \inparg{\leftskip = 40pt\rightskip = 40pt}
\def \outparg{\leftskip = 0 pt\rightskip = 0pt}
\def\lf{16\pi^2}
\def\beqn{\begin{eqnarray}}
\def\eeqn{\end{eqnarray}}
\def\sy{supersymmetry}
\def\sic{supersymmetric}
\def\sa{supergravity}
\def\ssm{supersymmetric standard model}
\def\sm{standard model}
\def\ssb{spontaneous symmetry breaking}
\def\smgroup{$SU_3\otimes\ SU_2\otimes\ U_1$}
\def\app{{Acta Phys.\ Pol.\ }{\bf B}}
\def\anp{Ann.\ Phys.\ }
\def\cmp{Comm.\ Math.\ Phys.\ }
\def\fortphys{{Fort.\ Phys.\ }{\bf A}}
\def\ijmpa{{Int.\ J.\ Mod.\ Phys.\ }{\bf A}}
\def\jetp{JETP\ }
\def\jetpl{JETP Lett.\ }
\def\jmp{J.\ Math.\ Phys.\ }
\def\mpla{{Mod.\ Phys.\ Lett.\ }{\bf A}}
\def\nc{Nuovo Cimento\ }
\def\npb{{Nucl.\ Phys.\ }{\bf B}}
\def\physrep{Phys.\ Reports\ }
\def\plb{{Phys.\ Lett.\ }{\bf B}}
\def\pnas{Proc.\ Natl.\ Acad.\ Sci.\ (U.S.)\ }
\def\pr{Phys.\ Rev.\ }
\def\prd{{Phys.\ Rev.\ }{\bf D}}
\def\prl{Phys.\ Rev.\ Lett.\ }
\def\ptp{Prog.\ Th.\ Phys.\ }
\def\sjnp{Sov.\ J.\ Nucl.\ Phys.\ }
\def\tmp{Theor.\ Math.\ Phys.\ }
\def\pw{Part.\ World\ }
\def\zpc{Z.\ Phys.\ {\bf C}}
\def\pa{\partial}

\vskip .3in {\textbf{\Large{ Non-relativistic approximate numerical ideal-magneto hydrodynamics of (1+1) D transverse flow in Bjorken scenario}}}
\medskip
\vskip .3in \centerline{ \bf  M. Haddadi Moghaddam$^{1,3}$,  B. Azadegan   $^1$,  A. F. Kord$^{1,2}$, W. M. Alberico$^3$ }
\bigskip
{\small{\it {$^1$: Department of Physics, Hakim Sabzevari University (HSU), P.O.Box 397, Sabzevar, Iran\\
 $^2$: Institute for Research  in Fundamental Sciences(IPM),P.O. Box 19395-5531, Tehran, Iran\\
$^3$: Department of Physics, University of Turin and INFN, Turin, Via P. Giuria 1, I-10125 Turin, Italy }}}


\small{\it\center{E-mail: afarzaneh@hsu.ac.ir}} \vskip .3in

\begin{abstract}
In this study,  we investigate  the impact of the magnetic field on the evolution of the transverse flow
of QGP matter in the magneto-hydrodynamic (MHD) framework. We assume that the magnetic field is
perpendicular to the reaction plane and then we solve the coupled Maxwell and conservation equations
in (1+1D) transverse flow, within the Bjorken scenario.  We consider a QGP with infinite electrical conductivity. First, the magnetic effects on the QGP medium  at mid-rapidity are investigated at leading order; then the  time and space dependence of
the energy density, velocity and magnetic field in the  transverse plane of
the ideal magnetized hot plasma are obtained.
 \end{abstract}

\section{Introduction}
 It is commonly accepted nowadays that  collisions of relativistic heavy-ions create
 a hot and dense fireball matter. Quarks and gluons are in a deconfined state, called quark gluon plasma(QGP),
 for a very short time ($\sim$ 1fm/c) after the initial hard parton collisions of nuclei.
The hydrodynamic approach has given one of the best description for the QGP matter: especially for
estimating the lowest shear viscosity over the entropy ratio,  this theoretical framework has shown
acceptable consistencies with many experimental results \cite{Romatschke:2007mq}-\cite{DelZanna:2013eua}.

 Recently, it has been  shown that in the peripheral AA-collisions such as Pb-Pb at the center of
mass energy $ \sqrt{s}=2.76$ TeV and Au-Au at the center of mass energy $\sqrt{ s }= 200$ GeV a
huge magnetic field is created, of the order of eB$\sim 10^{18}-10^{19}$ G,
which is $10^{13}$ times larger than the strongest steady magnetic field ever realized in the laboratory.
It has been claimed that the existence of such strong fields may be important for a variety of
new phenomena like  the Chiral Magnetic Effect (CME), Chiral Magnetic Wave (CMW),
Chiral Electric Separation Effect (CESE), Chiral Hall Separation Effect (CHSE),
pressure anisotropy in QGP, influence on the direct and elliptic flow,
shift of the critical temperature. A series of reviews and more references can be found
in Refs.~\cite{Kharzeev:2007jp}-\cite{a11}.
Hence, it will be worth to further investigate the properties of the QGP in the presence of EM fields.

There have been several works, which have  explored the  behavior of the space-time evolution
of electromagnetic fields created by the colliding charged beams moving at relativistic speed
in $z$-direction, as a solution of the Maxwell equations \cite{Tuchin:2013apa}-\cite{chinees}.
In ref \cite{Tuchin:2013apa}-\cite{Deng:2012pc}, for the sake of simplicity,  the  classical
Ohm law for induced currents in QGP has been suggested.  In many studies, it has been assumed that
there are no couplings between electrodynamic   and hydrodynamic  equations in a QGP medium.
Based on this assumption, it has been shown that the electromagnetic field  depends only on
the impact parameter of the colliding nucleons $b$, on the center of mass energy $\sqrt{s}$
and on the electric and chiral magnetic conductivities of the QGP;  besides, its decrease with time
is much slower than in vacuum.
In addition,  in several works (see, e.g., Refs.  \cite{Tuchin:2013apa}-\cite{chinees}) the
electromagnetic field is derived from Maxwell equations without coupling
to the velocity of the fluid, assuming that the latter has negligible
influence on the field itself.

 According to this point, considering the Bjorken flow four velocity,
 the electromagnetic response of QGP  in a quantum regime has been investigated in ref.~\cite{a20},
 and it has been concluded that the induced electric current in the plasma fireball
 cannot generate a classical electromagnetic field.
 In \cite{Gursoy:2014aka} charged dependence of flow coefficients has been discussed,
and the  effects of the EM field on  directed flow has been studied, showing that these effects are negligible.
However, we claim that this controversial
result has been  obtained by imposing that the velocity of charged particles $\vec v$ is
smaller than the velocity of the expanding plasma $\vec u$.

 Other works obtained a series of preliminary results, by estimating
the significance of strong EM fields on the QGP medium \cite{Bzdak:2011yy}-\cite{a27}.
In  most of them, it has been assumed the Maxwell equations
decouple from the time evolution of the QGP, and then
the evolution of the EM fields as well as their influence  on the flow coefficients have been studied.
Results have revealed that after collision, the strength of the EM fields  decrease.
In addition, it was found that  the ratio of magnetic pressure over the thermal pressure
$b^2/P$ of the hot in-viscid fluid is negligible.
However, the presence of a medium with {\it finite} electrical conductivity can substantially delay
the decay of the magnetic field~\cite{a27}.

It is obvious that the resulting EM field  is a solution of a complicated magneto-hydrodynamic
problem~\cite{Romatschke:2007mq}-\cite{DelZanna:2013eua}.
In fact, the relativistic magneto-hydrodynamic (RMHD) setup is one of the necessary tools in order
to describe the hot plasma in the presence of EM fields~\cite{a29}-\cite{an89}.
For this purpose, one needs a numerical code that solves the equations of (1+3) dimensional
relativistic magneto-hydrodynamics (RMHD).

 Recently, in refs.~\cite{roy15}-\cite{pu} some efforts have been made toward both numerical and analytical
 approaches aimed to solve the RMHD setup, by considering some constraints, specific of high energy heavy
 ion collisions. In ref.~\cite{pu} the main goal was to obtain an analytical solution for a (1+1) dimensional
Bjorken flow within ideal transverse RMHD; these authors have neglected (consistently with their hypothesis)
the coupling to Maxwell’s equations and have analytically solved the energy-momentum conservation equations
in a perturbation framework.

Another recent work employs a (1+3) dimensional RMHD code~\cite{Inghirami etal}: these authors
have used the initial conditions according to the solutions obtained from Maxwell equations
in the early-time of the collision: there are, however, many uncertainties
in the conditions of the pre-equilibrium phase.

In this paper we improve previous researches by removing some of the above mentioned restrictions: in
particular
we simulate (1+1) dimensional ideal magneto-hydrodynamics in the Bjorken  scenario
 to determine the effect of the  magnetic field on the behavior of an inviscid fluid.
Here, we consider the combination of relativistic hydrodynamic  equations with Maxwell equations and
solve numerically in (1+1) dimensions a set of {\it coupled} MHD equations.
 This improves some previous, analogous, work, where the coupling between Maxwell equations and conservation
equations has been neglected or treated perturbatively.
For the purpose of numerical calculations, we have supplemented a relatively simple code which
incorporates the contribution of a coupled electromagnetic field in (1+1) dimensions.
One important novelty is that we use the  boundary conditions at late time
($\tau \rightarrow \infty$): indeed the late-time dynamics has been governed by ideal hydrodynamics
and is known, while the early time conditions are unknown.
In order to check our code, we compare  our results with the analytical solutions of ref.~\cite{pu}.
We  find, indeed, that their results can be recovered by the numerical solutions,
at least in (1+1)-D transverse evolution.

The paper is organized as follows: in Section $2$,  we introduce the ideal relativistic magneto-hydrodynamic
equations in their most general form, considering them in the case of a plasma with infinite electrical conductivity.
In  Section $3$  we present  our numerical procedure with details in the setup; results obtained
with the spatial initial condition are shown in Section~4.
Finally, we summarize our conclusions and possible subsequent outlook in the last Section.

\section{Ideal transverse MHD setup in (1+1)D expansion}

We consider the relativistic magnetohydrodynamic (RMHD)  framework, in order to describe the
interaction of matter and electromagnetic fields in quark-gluon  plasmas \cite{a29}-\cite{an89}.
For the sake of simplicity, we assume an ideal
relativistic plasma  with  massless particles and  infinite electrical conductivity.
In addition, the fluid is considered  to be ultra relativistic, thus implying that the rest mass
contributions to the equation of state (EOS) have been  neglected, and the pressure is simply
proportional to the energy density: $P = c_s^2\epsilon=\frac{1}{3}\epsilon$ where $c_s=\sqrt{\frac{1}{3}}$
is the speed of sound.
 For an ideal fluid with infinite electrical conductivity, the equations of RMHD can
be written in the form of the covariant conservation laws
\begin{eqnarray}
d_\mu T^{\mu\nu}&=&0,\label{1}\\
d_\mu F^{\star\mu\nu}&=&0\label{2}
\end{eqnarray}
where
\begin{eqnarray}
T^{\mu\nu}&=&T_{matter}^{\mu\nu}+T_{EM}^{\mu\nu},\\
T_{matter}^{\mu\nu}&=&(\epsilon+P)u^\mu u^\nu+Pg^{\mu \nu}\\
T_{EM}^{\mu\nu}&=&b^2u^\mu u^\nu+\frac{1}{2}b^2g^{\mu\nu}-b^\mu b^\nu\\
F^{\star\mu\nu}&=&u^\mu b^\nu-u^\nu b^\mu,
\end{eqnarray}
and
\begin{eqnarray}
b^\mu=F^{\star\mu\nu}u_\nu, \ \  (b^{\mu}u_{\mu}=0),\ \ b^2=b^\mu b_\mu
\end{eqnarray}
  Here $ F^{\star\mu\nu}$ is the dual tensor of electromagnetic field.
$\epsilon$ and $P$ are energy density and pressure respectively. $b^{\mu}$ is the  magnetic field four
vector in the local rest-frame of the  fluid, which is related  in the standard way to the one measured in the lab-frame.
In the present paper we assume a  fluid with infinite electrical conductivity,
so  the  electric  field four vector in the local rest-frame equals to zero ($e^{\mu}=0$).
 Besides,  the single fluid  four velocity $u_{\mu}$ ($u_{\mu}u^{\mu}=-1$) is defined  as follow:
$$u_\mu=\gamma(1, \vec v),\ \gamma=\frac{1}{\sqrt{1-v^2}}$$
In eqs.(\ref{1}) and (\ref{2}) the covariant derivative is given by:
\begin{eqnarray}
d_\mu A^\nu&=&\partial_\mu A^\nu+\Gamma^\nu_{\mu m} A^m\\
d_p A^{\mu\nu}&=&\partial_p A^{\mu\nu}+\Gamma^\mu_{p m} A^{m \nu}+\Gamma^\nu_{p m} A^{m \mu},
\end{eqnarray}
where $\Gamma^i_{j k}$ are the Christoffel symbols
\begin{eqnarray}
\Gamma^i_{jk}=\frac{1}{2}g^{im}\left(\frac{\partial g_{mj}}{\partial x^k}+\frac{\partial g_{mk}}
{\partial x^j}-\frac{\partial g_{jk}}{\partial x^m}\right)
\end{eqnarray}
 and $g_{ij}$ the metric tensor.

It is more convenient to work with Milne coordinates rather than the standard Cartesian coordinates
for a longitudinally boost-invariant flow:
\begin{eqnarray}
(\tau, x, y, \eta)&=&\left(\sqrt{t^2-z^2},x,y,\frac{1}{2}ln\frac{t+z}{t-z}\right).
\end{eqnarray}
Here, the metric is given by:
\begin{eqnarray}
g^{\mu\nu}=diag(-1, 1, 1, 1/\tau^2),  \  \  \  \ g_{\mu\nu}=diag(-1, 1, 1, \tau^2)
\end{eqnarray}

Working in  Milne coordinates, one can easily obtain the Christoffel symbols: the only non-zero ones
being: $\Gamma^\tau_{\eta\eta}=\tau$ and $\Gamma^\eta_{\tau\eta}=1/\tau$.
Then, four distinct  conservation equations  can be easily derived
from $d_\mu T^{\mu\nu}=0$ in the Milne coordinate system. They are given by:
\begin{eqnarray}
&&\partial_\tau T^{\tau\tau}+\partial_{x}T^{x \tau }+\partial_{y}T^{y\tau }+
\partial_{\eta}T^{ \eta\tau}+\tau T^{\eta\eta}+\frac{1}{\tau}T^{\tau\tau}=0,\label{C1}\\&&
\partial_\tau T^{\tau x}+\partial_{x}T^{x x}+\partial_{y}T^{y x}+\partial_{\eta}T^{\eta x}+
\frac{1}{\tau}T^{\tau x}=0,\label{C2}\\&&
\partial_\tau T^{\tau y}+\partial_{x}T^{x y}+\partial_{y}T^{y y}+\partial_{\eta}T^{\eta y}+
\frac{1}{\tau}T^{\tau y}=0,\label{C3}\\&&
\partial_\tau T^{\tau\eta}+\partial_{x}T^{x \eta}+\partial_{y}T^{y \eta}+\partial_{\eta}T^{\eta \eta}+
\frac{3}{\tau}T^{\tau \eta}=0\label{C4}
\end{eqnarray}
In contrast with the energy momentum tensor $T^{\mu\nu}$, the dual electromagnetic tensor $F^{*\mu\nu}$  is
antisymmetric; hence the homogeneous Maxwell equation, $d_\mu F^{*\mu\nu}=0$,
 leads to the following equations:
\begin{eqnarray}
\partial_x F^{*x\tau}+\partial_y F^{*y\tau}+\partial_\eta F^{*\eta\tau}&=&0,\label{M1}\\
\partial_\tau F^{*\tau x}+\partial_y F^{*y x}+\partial_\eta F^{*\eta x}+\frac{1}{\tau}F^{*\tau x}&=&0,\label{M2}\\
\partial_\tau F^{*\tau y}+\partial_x F^{*x y}+\partial_\eta F^{*\eta y}+\frac{1}{\tau}F^{*\tau y}&=&0,\label{M3}\\
\partial_\tau F^{*\tau \eta}+\partial_x F^{*x\eta}+\partial_y F^{*y \eta}+\frac{1}{\tau}F^{*\tau\eta}&=&0.\label{M4}
\end{eqnarray}

In order to simplify the  problem,  we assume that
the  magnetic field is perpendicular to the reaction plane, pointing along the $y$
direction in an inviscid fluid with infinite electrical conductivity,
 following the Bjorken expansion along the $z$ direction and moving, in the transverse plane, only in $x$ direction.
The boost invariance of the Bjorken expansion allows us to restrict the discussion to the $z=0$ plane, where symmetry
reasons impose $u^z=0$.
Then:
\begin{eqnarray}
u_\mu=\tilde\gamma(1, u_x, 0, 0),\ b_\mu=(0, 0, b_y, 0),\ e_\mu=(0, 0, 0, 0).
\end{eqnarray}
Where $\tilde\gamma=\frac{1}{\sqrt{1-u_x^2}}$, and $u^\mu b_\mu=0,\ u^\mu u_\mu=-1$ are satisfied.

In our setup,  the energy momentum and  dual electromagnetic tensors  are given by:
\begin{eqnarray}
T^{\mu\nu}=\begin{pmatrix} (\epsilon+P+b^2)\tilde\gamma^2-P-\frac{b^2}{2} & -(\epsilon+P+b^2)\tilde\gamma^2 u_x &0 &0  \\
-(\epsilon+P+b^2)\tilde\gamma^2 u_x & (\epsilon+P+b^2)\tilde\gamma^2 u_x^2+P+\frac{b^2}{2} & 0 & 0\\
0  & 0 & P-\frac{b^2}{2} & 0 \\
 0& 0 & 0&( P+\frac{b^2}{2})(\frac{1}{\tau^2})\end{pmatrix}\label{Tmn}
 \end{eqnarray}

\begin{eqnarray}
F^{*\mu\nu}&=\begin{pmatrix} 0 & 0 &-\tilde\gamma b_y &0  \\
0 & 0 & \tilde\gamma u_x b_y & 0\\
\tilde\gamma b_y  & -\tilde\gamma u_x b_y & 0 & 0 \\
 0& 0 & 0& 0\end{pmatrix}\label{Fmn}
\end{eqnarray}

When the  Eqs.(\ref{Tmn})--(\ref{Fmn}) are plugged into Eqs.(\ref{C1})--(\ref{M4}), one obtains
the following coupled equations:

\begin{eqnarray}
&&\partial_\tau\left[(\epsilon+P+b^2)\tilde\gamma^2-P-\frac{b^2}{2}\right]+
\partial_x\left[-(\epsilon+P+b^2)\tilde\gamma^2 u_x\right]\nonumber\\&&
+\frac{(\epsilon+P+b^2)\tilde\gamma^2}{\tau}=0,\label{F1}\\&&
\partial_\tau\left[-(\epsilon+P+b^2)\tilde\gamma^2 u_x\right]
+\partial_x\left[(\epsilon+P+b^2)\tilde\gamma^2 u_x^2+P+\frac{b^2}{2}\right]\nonumber\\&&
-\frac{(\epsilon+P+b^2)\tilde\gamma^2 u_x}{\tau}=0,\label{F2}\\&&
\partial_y\left(P-\frac{b^2}{2}\right)=0,\label{F3}\\&&
\partial_\eta\left(( P+\frac{b^2}{2})(\frac{1}{\tau^2})\right)=0,\label{F4}\\&&
\partial_y(\tilde\gamma b_y)=0,\label{F5}\\&&
\partial_y(\tilde\gamma u_x b_y)=0.\label{F6}\\&&
\left[-\tilde\gamma\partial_\tau+\tilde\gamma u_x \partial_x\right]b_y+b_y
\left(\partial_x(\tilde\gamma u_x)-\partial_\tau(\tilde\gamma)-\frac{\tilde\gamma}{\tau}\right)=0
\label{F7}
\end{eqnarray}

In the following, we will assume  that the transverse velocity $u_x$ is  non-relativistic,
so we will keep only first order terms in $u_x$.
Since we consider (1+1)D flow,  all thermodynamical variables depend only on  $\tau, x$ coordinates.
 Applying  the definition of $u^2=-1$ one  finds that $u_\mu=(1, u_x, 0, 0)$ and
$\tilde\gamma\rightarrow 1$.
Using all the above assumptions the set of equations (\ref{F1})--(\ref{F7}) reduce to:
\begin{eqnarray}\label{eq:inductionfinal1}
&&(-\partial_\tau+u_x\partial_x) b_y+(\frac{\partial u_x}{\partial x}-\frac{1}{\tau}) b_y=0,\\
\label{eq:inductionfinal2} &&-\partial_\tau\epsilon+u_x\partial_x\epsilon+(1+c_s^2)\epsilon
(\frac{\partial u_x}{\partial x}-\frac{1}{\tau})=0,\\
\label{eq:inductionfinal3}
&&-\Big((1+c_s^2)\epsilon+b^2\Big)\partial_\tau u_x+c_s^2\partial_x \epsilon+b_y
\partial_x b_y+\left(\frac{c_s^2(1+c_s^2)\epsilon+b^2}{\tau}\right)u_x =0.
\end{eqnarray}

As one expects in the ideal MHD, the energy conservation equation (\ref{eq:inductionfinal1} does
not include the $B$ field.

In the next section we present a numerical method  to solve the above coupled equations
(\ref{eq:inductionfinal1})--(\ref{eq:inductionfinal3}) simultaneously.


\section{Numerical calculation}
In this section we will solve the coupled relativistic hydrodynamic and Maxwell equations, which are
summarized in Eqs.(\ref{eq:inductionfinal1})--(\ref{eq:inductionfinal3}).
The solutions of the three coupled differential equations will be obtained by using the numerical
method of lines (MOL). This method is a technique for solving partial differential equations (PDE)
by discretizing one variable in one of the two dimensions
and then by integrating the semi-discrete problem as a system of ordinary differential equations (ODE).
Here we discretize the partial derivatives with respect to the space variables and obtain a system of
ODEs in the time variable: then the initial value software Mathematica has been used to solve
this ODE system. It is necessary that the partial differential equation problem be well posed as
an initial value problem in at least one dimension, since these are the conditions for an appropriate
use of the employed ODEs integrators.

Hence we discretize the coordinate $x$ with $N$ ($N$ even) uniformly spaced grid points
$x_i=(i-1)h,\ x_{N+1}=\pi,\ i=1,2,…,N$ and $h=\pi/N$.
We use a second-order finite difference formula for the first derivative in $x$.
In this configuration $v_i(\tau)$ indicates $v(\tau, x_i)$. In Fig. 1, the lines along which the
discrete quantities $v_i(\tau)$ are defined, are shown. Using the second-order difference
approximation for the first derivative in $x$ results in
\begin{eqnarray}
\frac{dv_i(\tau)}{dx}=\frac{-3v_i(\tau)+4v_{i+1}(\tau)-v_{i+2}(\tau)}{2h},\ i=1,...., N+1\,.
\end{eqnarray}
After substituting the first derivatives with respect to $x$ for the $v_i(\tau),\ \epsilon_i(\tau),\ b_i(\tau)$,
in equations (\ref{eq:inductionfinal1})--(\ref{eq:inductionfinal3}) one is left with a set of coupled ODEs: in order
to numerically solve these equations the crucial point remains the definition of the boundary conditions.

\begin{figure}[ht]
\center{\includegraphics[width=.6\textwidth]{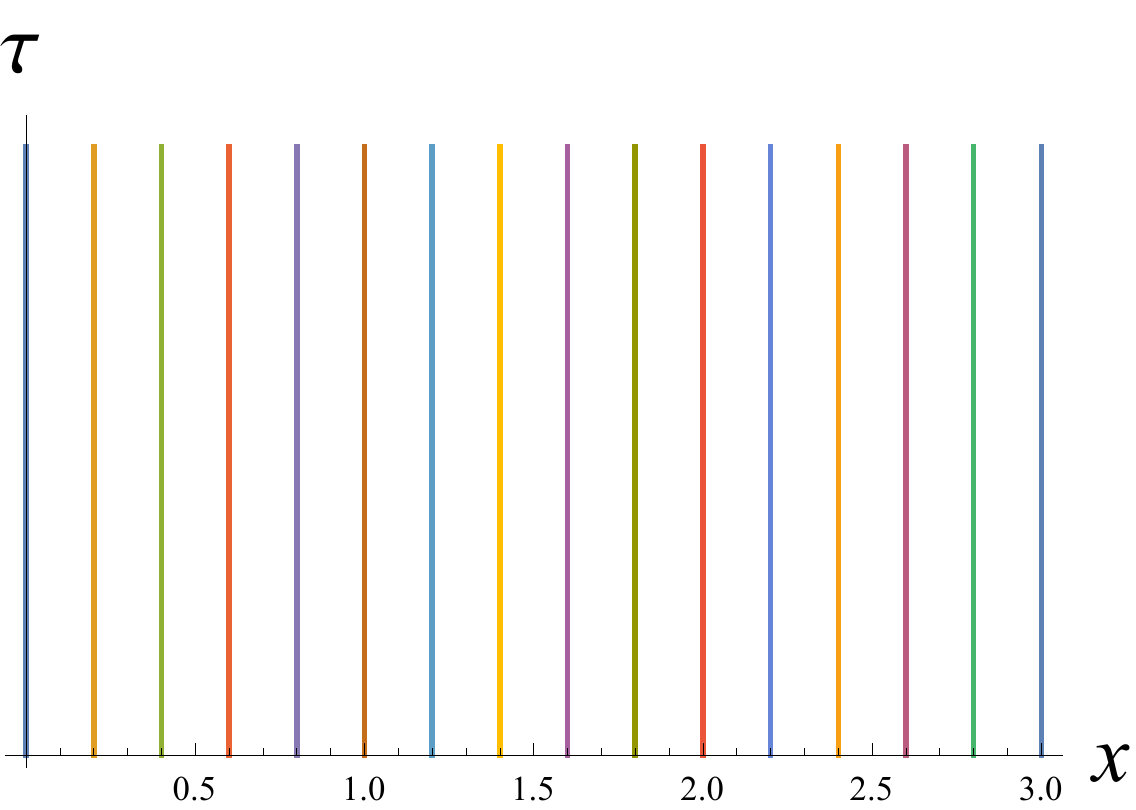}
\caption{The geometry of the PDEs in the MOL. The $v_i(\tau)$ is defined along the lines.}
\label{1selfsim_plot}}
\end{figure}
\begin{figure}[ht]
\center{\includegraphics[width=.8\textwidth]{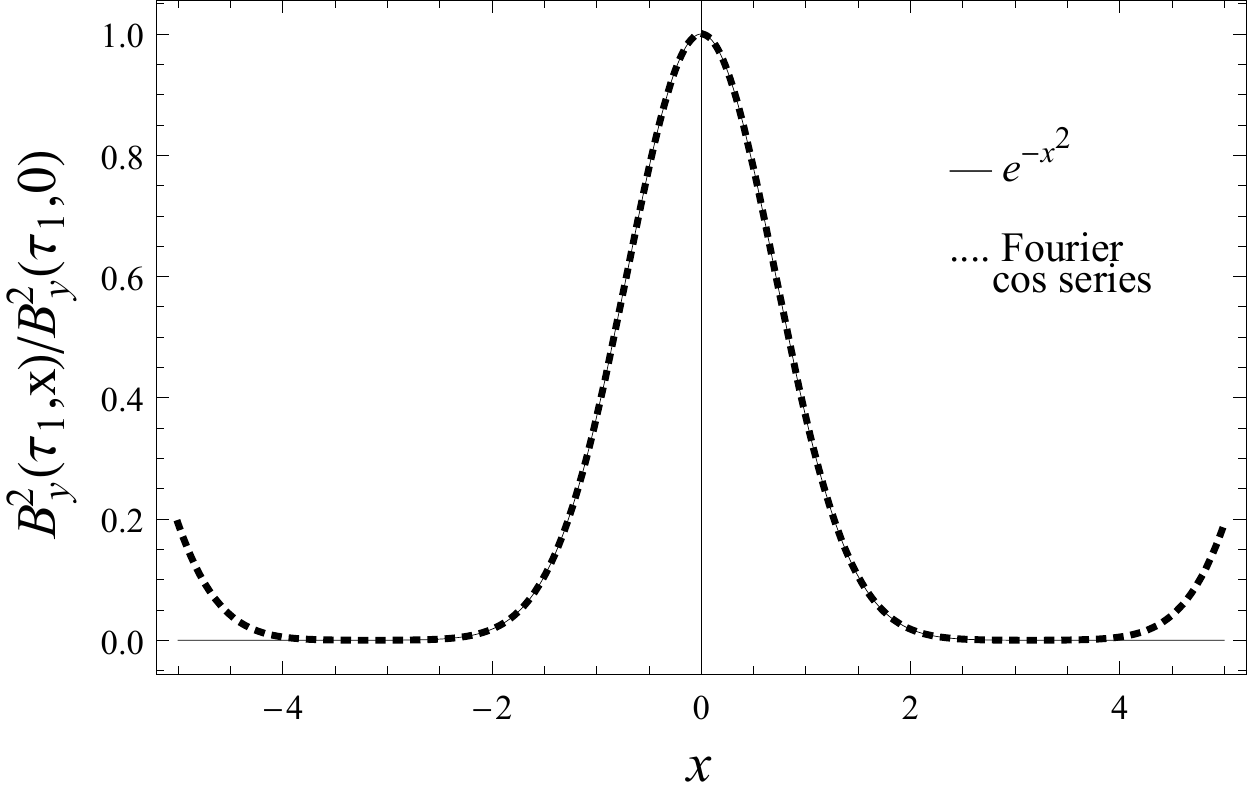}
\caption{A comparison between the approximated $b_y^2$
in Fourier cosine series (dots) and the Gaussian
distribution of $b_y^2$ (thin line).}
\label{2selfsim_plot}}
\end{figure}

\begin{figure}[ht]
\center{\includegraphics[width=.6\textwidth]{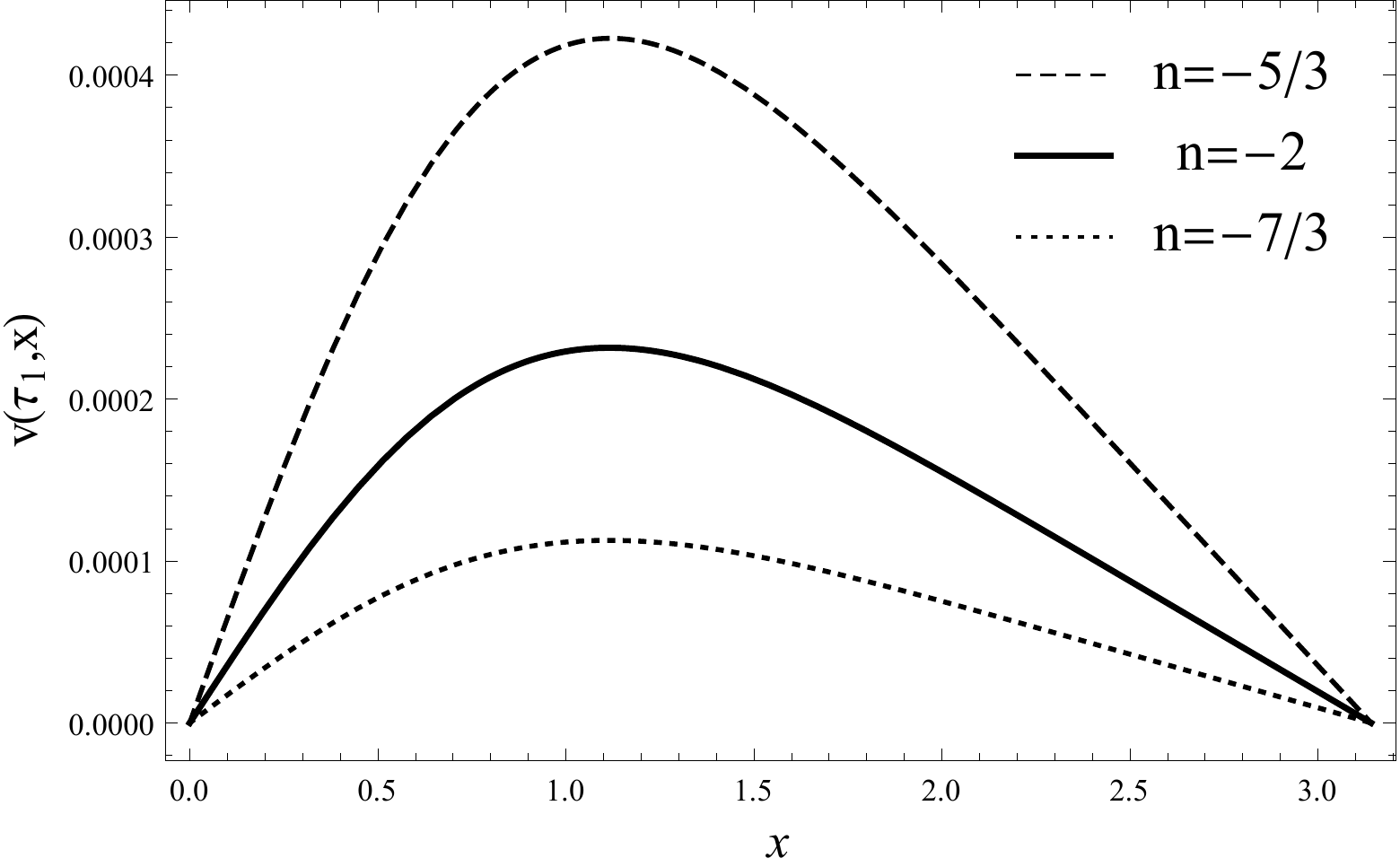}
\caption{Transverse velocity $v(\tau, x)$ versus $x$ plotted at the late time
$\tau_1=20$ fm with different value of $n$.}
\label{3selfsim_plot}}
\end{figure}
\begin{figure}[ht]
\center{\includegraphics[width=.6\textwidth]{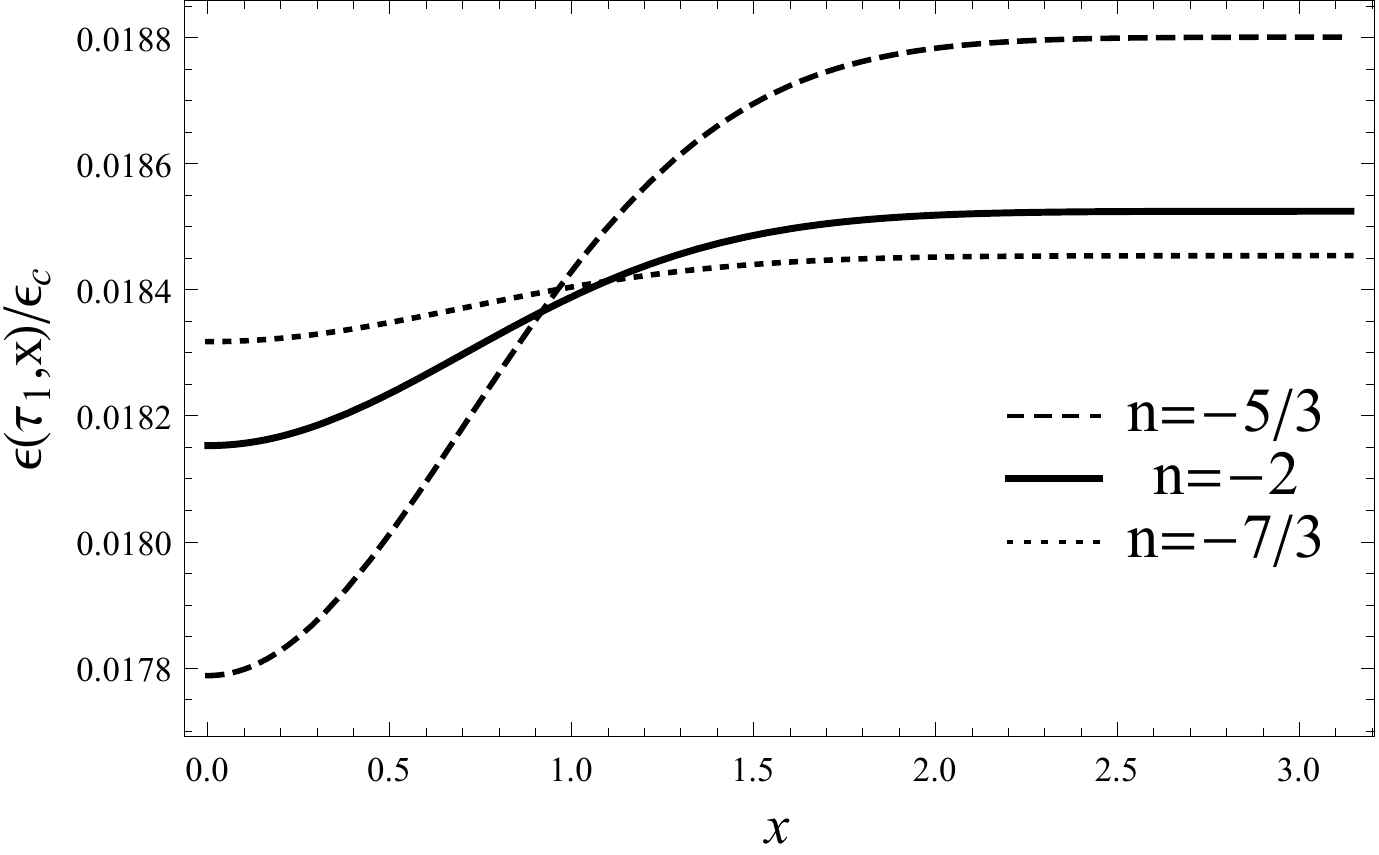}
\caption{Energy density $\epsilon(\tau, x)/\epsilon_c$ versus x plotted at the late time
$\tau_1=20$ fm with different value of $n$.}
\label{4selfsim_plot}}
\end{figure}

\begin{figure}[ht]
\center{\includegraphics[width=.6\textwidth]{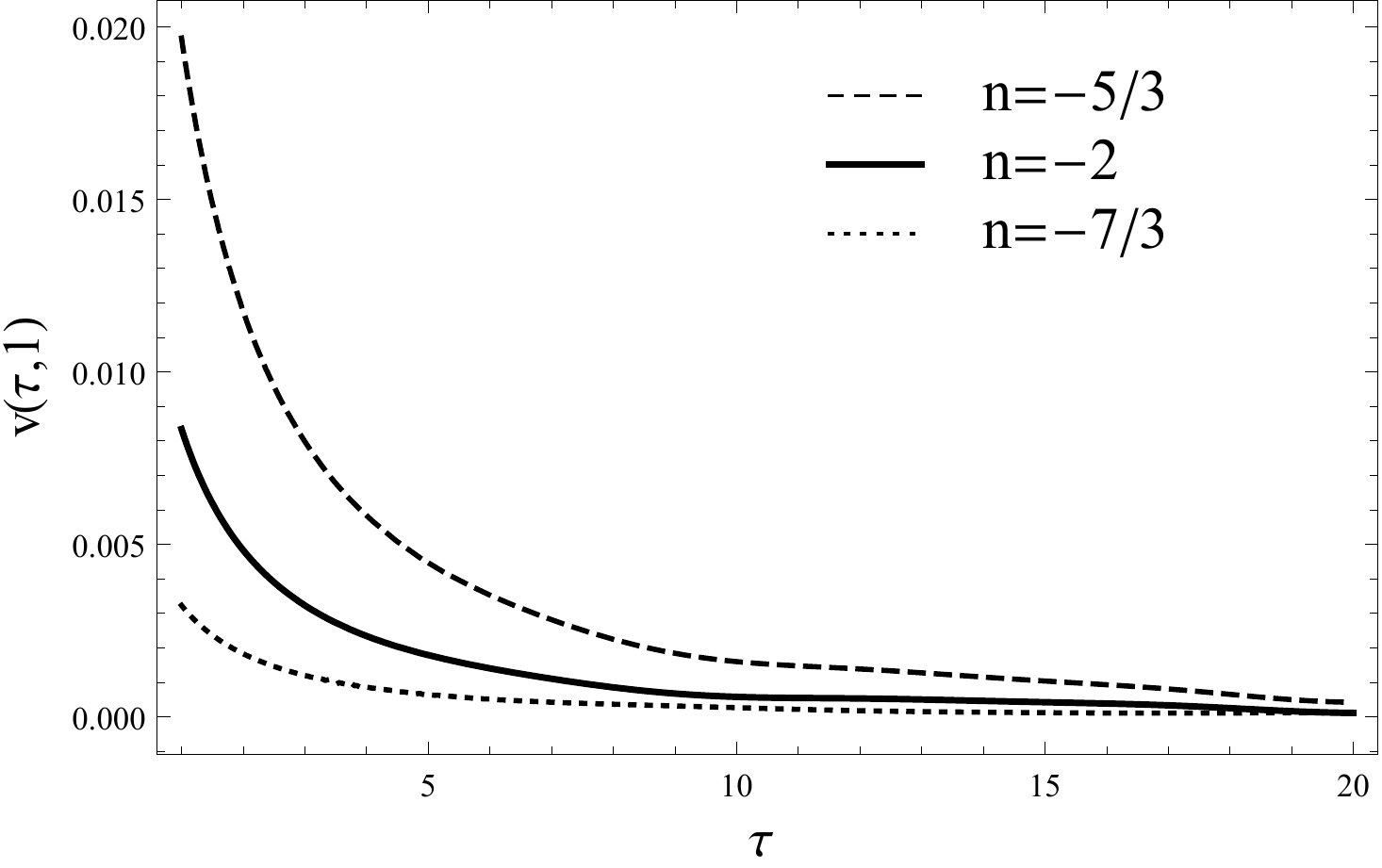}
\caption{Transverse velocity $v(\tau, x)$ versus $\tau$ plotted at $x=1$ with different
values of $n$. The dashed, solid and
dotted curves correspond to $ n=-5/3, -2$ and $-7/3$ respectively.}
\label{5selfsim_plot}}
\end{figure}
\begin{figure}[ht]
\center{\includegraphics[width=.6\textwidth]{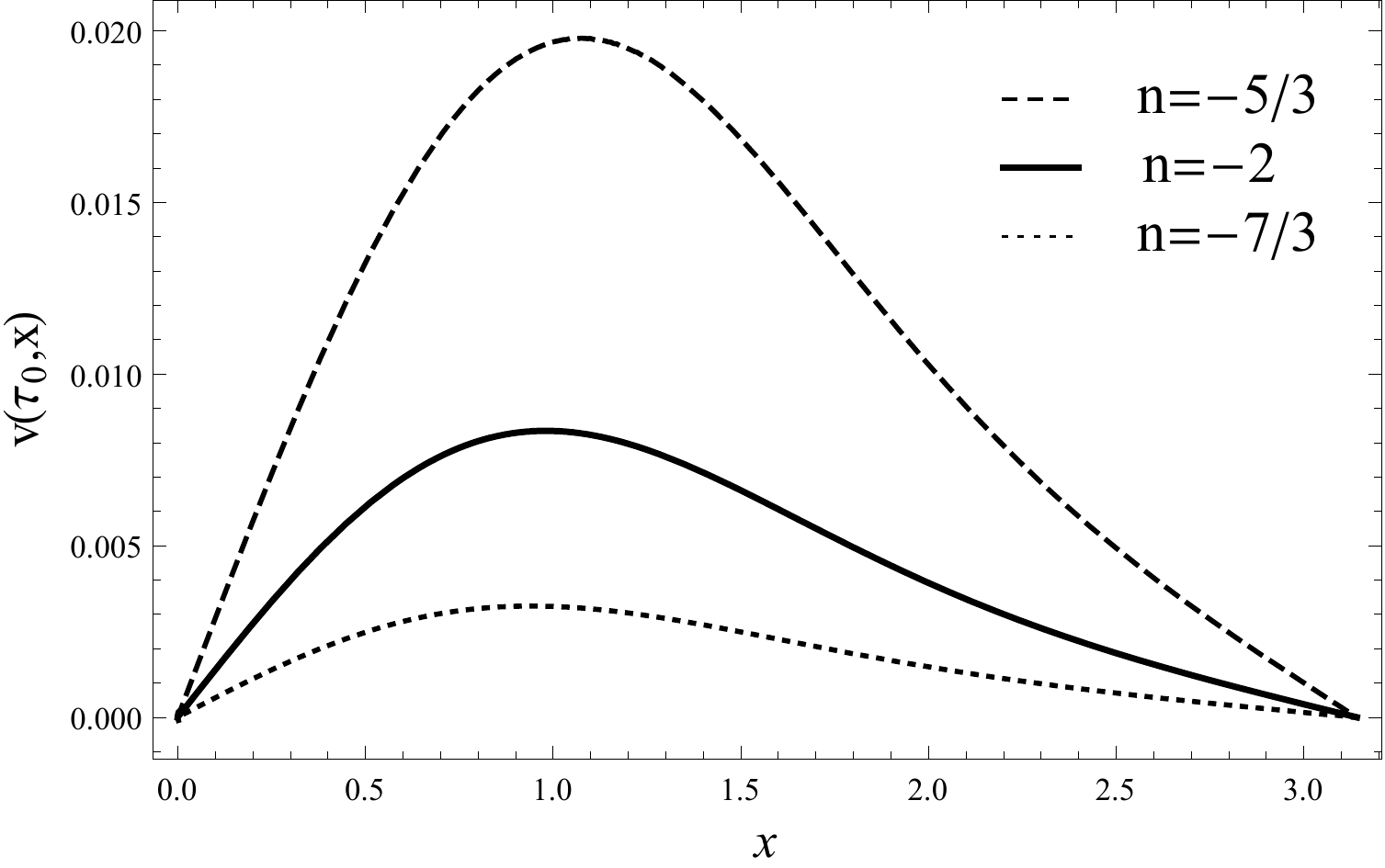}
\caption{Transverse velocity $v(\tau, x)$ versus $x$ plotted at $\tau=1$ fm with different values
of $n$. The dashed, solid and dotted curves correspond to $n=-5/3, -2$ and $-7/3$ respectively}
\label{6selfsim_plot}}
\end{figure}
\begin{figure}[ht]
\center{\includegraphics[width=.6\textwidth]{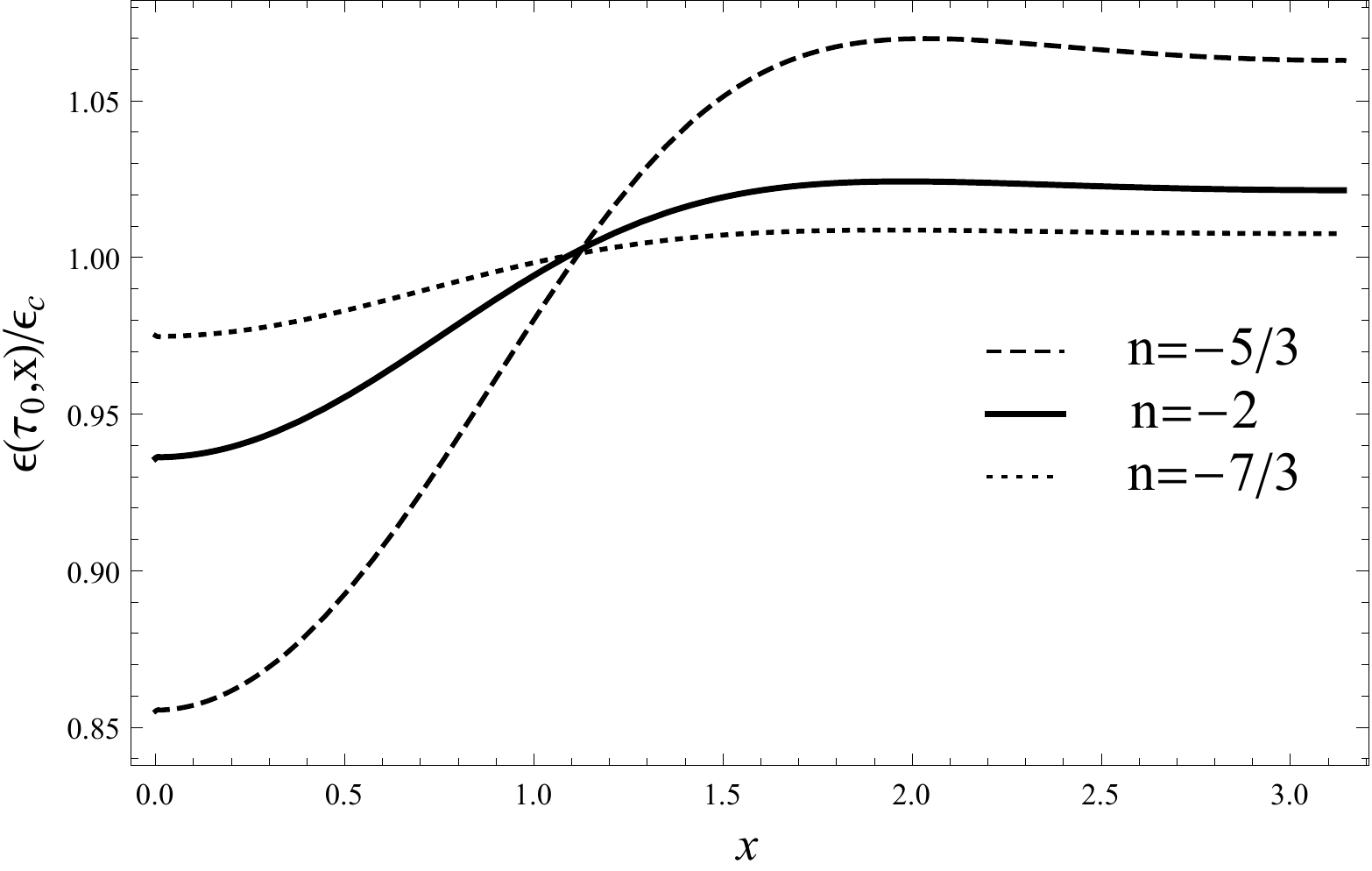}
\caption{Energy density $\epsilon(\tau, x)/\epsilon_c$ versus x plot at  $\tau_1=1$ fm
with different values of $n$. The dashed, solid and
dotted curves correspond to $ n=-5/3, -2$ and $-7/3$ respectively.}
\label{7selfsim_plot}}
\end{figure}

\begin{figure}[ht]
\center{\includegraphics[width=.6\textwidth]{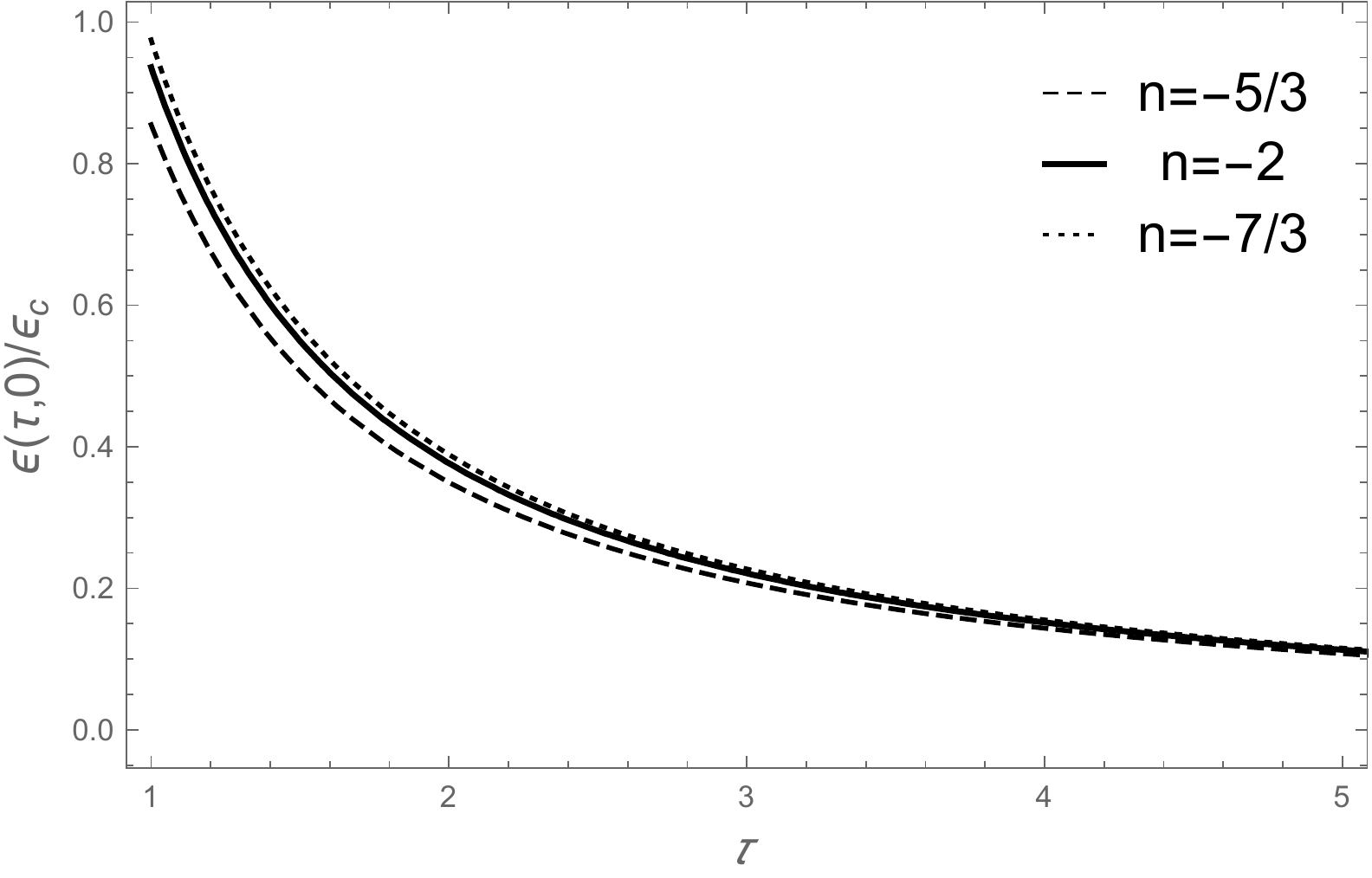}
\caption{Energy density $\epsilon(\tau, x)/\epsilon_c$ versus $\tau$ plot at  $x=0$ fm with different values of $n$. The dashed, solid and
dotted curves correspond to $ n=-5/3, -2$ and $-7/3$ respectively.}
\label{8selfsim_plot}}
\end{figure}

\begin{figure}[ht]
\center{\includegraphics[width=.6\textwidth]{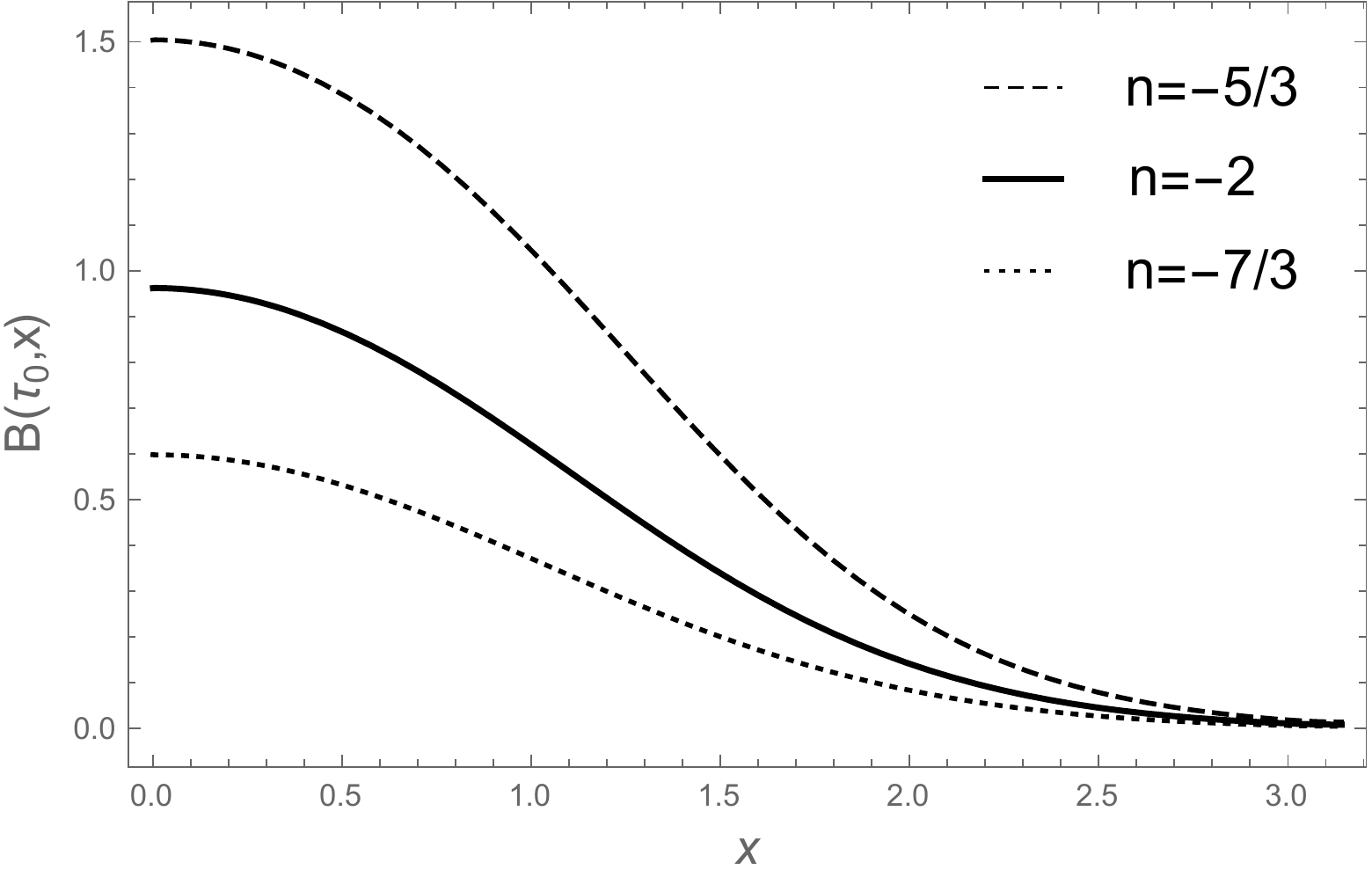}
\caption{Magnetic field $b_y(\tau, x)$ versus $x$ plot at $\tau=1$ with different values of $n$.
The dashed, solid and
dotted curves correspond to $ n=-5/3, -2$ and $-7/3$ respectively.}
\label{9selfsim_plot}}
\end{figure}

\begin{figure}[ht]
\center{\includegraphics[width=.6\textwidth]{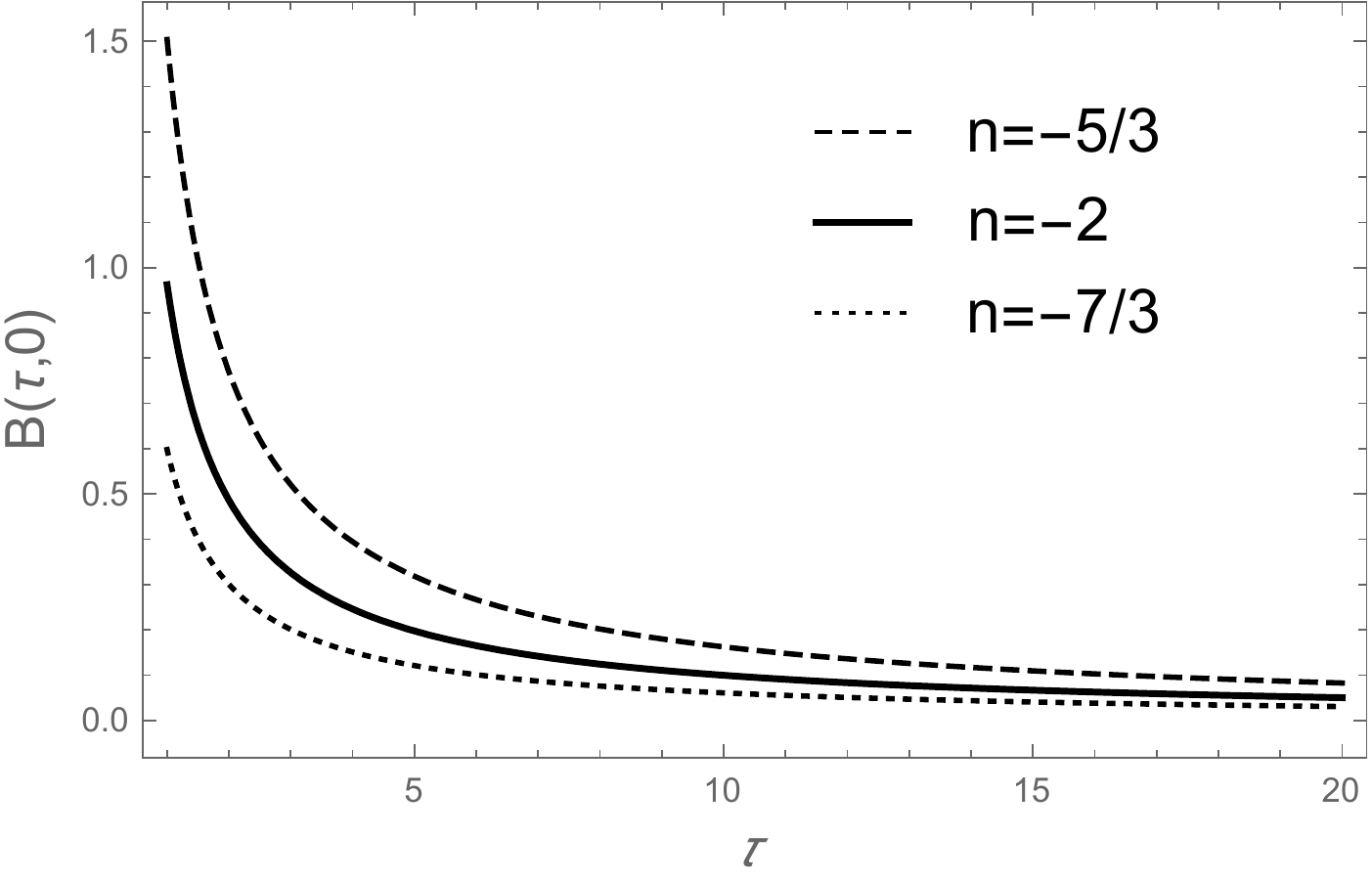}
\caption{Magnetic field $b_y(\tau, x)$ versus  $\tau$ plot at  $x=0$ fm with different values
of $n$.  The dashed, solid and dotted curves correspond to $ n=-5/3, -2$ and $-7/3$ respectively.}
\label{10selfsim_plot}}
\end{figure}


\begin{figure}[ht]
\center{\includegraphics[width=.6\textwidth]{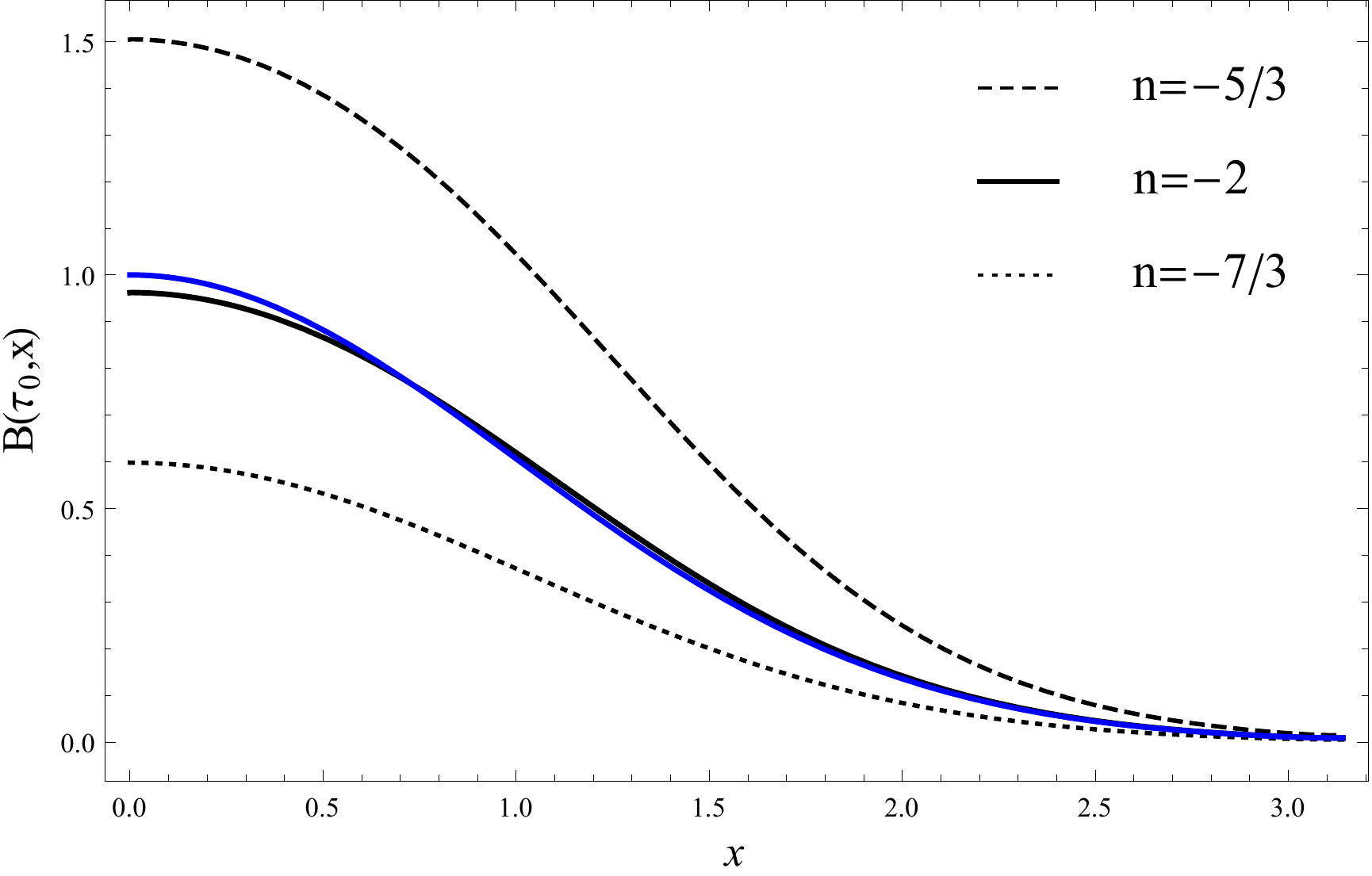}
\caption{Magnetic field $b_y(\tau, x)$ versus $x$ plot at $\tau=1$ with different values of $n$.
The black curves correspond to coupling of Maxwell’s equations with conservation equations
(present work) and the blue
curve correspond to the analytical solutions.}
\label{11selfsim_plot}}
\end{figure}


\subsection{Boundary conditions}
In our work we consider the  boundary conditions at late-time.
Indeed the issue under investigation does not allow to have precise information on the early-times
conditions. Also the determination of the proper   boundary conditions from  analytical solutions of
the equations(\ref{eq:inductionfinal1})--(\ref{eq:inductionfinal3}) at late-time ($\tau\to\infty$)
is still very difficult, due to the coupling between conservation and Maxwell equations.
Hence we  derived such  boundary conditions at late time from the analytical
solutions of ref\cite{pu}.These authors have investigated the magneto hydrodynamics
in the presence of an external magnetic field, which follows the power-law decay in proper time and
has spatial inhomogeneity characterized by a Gaussian distribution in transverse coordinates.
For simplicity they have  neglected the coupling of Maxwell’s equations and conservation equations and
solved the conservation equations perturbatively and analytically.

In ref.~\cite{pu} the profile of  the  magnetic field has been defined by:
\begin{eqnarray}
b_y(\tau, x)\hat y=B_c \tau^{n/2}e^{-x^2/2}\hat y
\label{magn}
\end{eqnarray}
Where $n$ is a negative value which governs the decay of magnetic field with increasing time.
The Fourier expansion of the   above square   magnetic field is  approximated as
\begin{eqnarray}
B_y(\tau, x)^2=B_c^2 \tau^n(0.28+0.44\cos x+0.21\cos 2x+0.06\cos 3x+0.01\cos 4x)
\end{eqnarray}
and Fig.~2 shows a comparison between the Fourier cosine series and the Gaussian distribution at the
late time $\tau_1=20$ fm.\footnote{In the ratio illustrated in Fig.~2 the $\tau$ dependence (and hence the precise
value of $\tau$)n is irrelevant, since it cancels in the ratio; however in the following this value will be
chosen to fix the initial late conditions.} Due to the oscillatory property of the cosine function, the
 solutions are
valid only in the region $-\pi<x<\pi$. The spatial width of the magnetic field depends on the impact parameter
of the considered peripheral collision.
Following the method which has been presented in  ref.~\cite{pu} one can obtain
analytical solutions   for transverse velocity $v(\tau ,x)$  and energy density ratio
$\epsilon(\tau, x)/\epsilon_c$ at the assumed late time [$\epsilon_c$ is the initial energy density
of the medium at time $\tau_0$].  They are  shown
in Figs.~3 and 4, for  $\tau_1=20$~fm and  3 different values of $n$; the value
$B_c^2/\epsilon_c=0.1$ has been assumed.

Then we  consider the following family of equations:
\begin{eqnarray}
\frac{\partial\epsilon_i(\tau)}{\partial\tau}&=&\frac{-3\epsilon_i(\tau)+4\epsilon_{i+1}
(\tau)-\epsilon_{i+2}(\tau)}{2h}v_i(\tau)+\frac{4}{3}\Big[\frac{-3v_i(\tau)+4v_{i+1}
(\tau)-v_{i+2}(\tau)}{2h}-\frac{1}{\tau}\Big]\epsilon_i(\tau),\nonumber\\
\frac{\partial b_i(\tau)}{\partial\tau}&=&\frac{-3b_i(\tau)+4b_{i+1}(\tau)-b_{i+2}(\tau)}{2h}
v_i(\tau)+\Big[\frac{-3v_i(\tau)+4v_{i+1}(\tau)-v_{i+2}(\tau)}{2h}-\frac{1}{\tau}\Big]
b_i(\tau),\nonumber\\
\frac{\partial v_i(\tau)}{\partial\tau}&=&\Big(\frac{4}{3}\epsilon_i(\tau)+b_i(\tau)^2\Big)^{-1}
\Big[\frac{1}{3}\Big(\frac{-3\epsilon_i(\tau)+4\epsilon_{i+1}(\tau)-b_{i+2}(\tau)}{2h}\Big)
 \nonumber\\
&&+b_i(\tau)\Big(\frac{-3b_i(\tau)+4b_{i+1}(\tau)-b_{i+2}(\tau)}{2h}\Big)+\Big(\frac{4}{9}
\epsilon_i(\tau)+b_i(\tau)^2\Big)\frac{v_i(\tau)}{\tau}\Big]
\end{eqnarray}

These are $3N+3$ first order, coupled ODEs with  boundary conditions:
$\epsilon_i(\tau_1)\equiv \epsilon_i(\tau_1,x_i), \ b_i(\tau_1)\equiv b_i(\tau_1,x_i),
\ v_i(\tau_1)\equiv v_i(\tau_1,x_i),  i=1, ..., N+1$ and $v_1(\tau)=\ v_{N+1}(\tau)=0$.
 These functions are  obtained from
analytical solutions of ref.~\cite{pu}   and shown in Figs. 2-4.

\section{Numerical results of MHD}

We shall now show the results obtained by numerically solving the above outlined system of equations.
To resume the procedure, we remind that with the method of lines we have fixed discrete values for
the variable x ($3N+3$ values) and defined derivatives with respect to $x$ via the second order difference
method; then the original set of equations reduces to a system of $3N+3$ coupled ordinary differential
equations for the quantities $\epsilon_i(\tau),\ b_i(\tau),\ v_i(\tau)$.

The above mentioned ODEs have been solved by using an ODE solver of Mathematica, with respect to the time variable.
The initial boundary conditions for the ODE integrator have been fixed at the late time $\tau_1=20$~fm and derived
from the analytical solutions of ref.~\cite{pu}.

This procedure allowed us to solve the (R)MHD equations in 1+1 dimensions
[Eqs.~(\ref{eq:inductionfinal1})--(\ref{eq:inductionfinal3})]
and to obtain the space time evolution of the magnetic field, the energy density  and the velocity of
plasma,  $v(\tau, x), \epsilon(\tau, x)$ and $b(\tau, x)$.
Our results for these functions are presented  in following figures, where they are plotted versus $ x$
at fixed $\tau$   or versus $\tau$ at  fixed $x$, for three different  values of $n$  ( $n<-1$).

 Figs.~5 and 6 show the variation of the fluid velocity in terms of either $\tau$ or $x$  with different values of $n$.
 Fig. 5 shows that  $|v|$  at fixed x is large at early times end becomes small in late times. This behavior is
 strikingly evident for the smallest $|n|$ employed here.
The transverse velocity $v$ in terms of $x$  at the fixed time $\tau=1$ fm has been plotted
 in Fig~6. $v$  increases from $x=0$  to a maximum  at intermediate x and gradually decrease with x.
 Besides, one can see that when the $|n|$  increases, the $v$ at fixed $\tau$ becomes smaller due to the
faster decay of the magnetic field.

 Figs.~7 and 8  show the energy density in terms of $x$ at fixed $\tau$  or in terms of $\tau$ at
 fixed $x$,   for three different  values of $n$  ( $n<-1$).  Fig.~7  indicates  that $\epsilon$ grows from $x=0$ up
 to some intermediate value of $x$, where it seems to saturate: the increase is more rapid when $n>-2$.
 The behavior of the energy density as a function of $\tau$ is monotonically decreasing;
 notice the different scale between Fig.~7 and Fig.~8.

Finally in Figs.~9 and 10 the magnetic field  $b_y(\tau, x)$ is plotted versus $x$  at fixed time ($\tau=1$) or
in terms of $\tau$ at  fixed $x$. Fig.~9 shows that $b_y(\tau, x)$ becomes small for large $x$. Besides,
when $|n|<2$ the magnetic field is quite larger, in the central area, than for the other values of n.

Next we wish to validate our numerical work by comparing it with the approximate analytical
calculation of Ref.~$\cite{pu}$: as already stated these authors assume an external, time-decreasing
magnetic field $B$ with a Gaussian distribution in $x$.

In Fig.~11. we show the profile of the magnetic field for different values of $n$, for the fixed time $\tau=1$:
the black curves correspond to our solutions and the blue curves correspond to the approximate analytical
solution of Ref.~$\cite{pu}$, where the magnetic field is considered as the one given in Eq.~(\ref{magn}).
At $\tau=1$ obviously the profiles of magnetic field Eq.~(\ref{magn}) is independent of the value of $n$,
while in the present work we obtain different results, as already seen in Fig.~9.
The case $n=-2$ corresponds to ideal magnetohydrodynamics and is referred to as the ''frozen-flux condition'',
which stems from Maxwell’s equations  with conservation of the entropy-density current: in this case the
analytical solution coincides with the one obtained in the present calculation.
Instead for $n=-7/3$ the magnetic field decreases by nearly a
factor 0.5 and for $n=-5/3$ it increases by a factor 1.5 relative to case $n=-2$.

\begin{figure}[ht]
\center{\includegraphics[width=.6\textwidth]{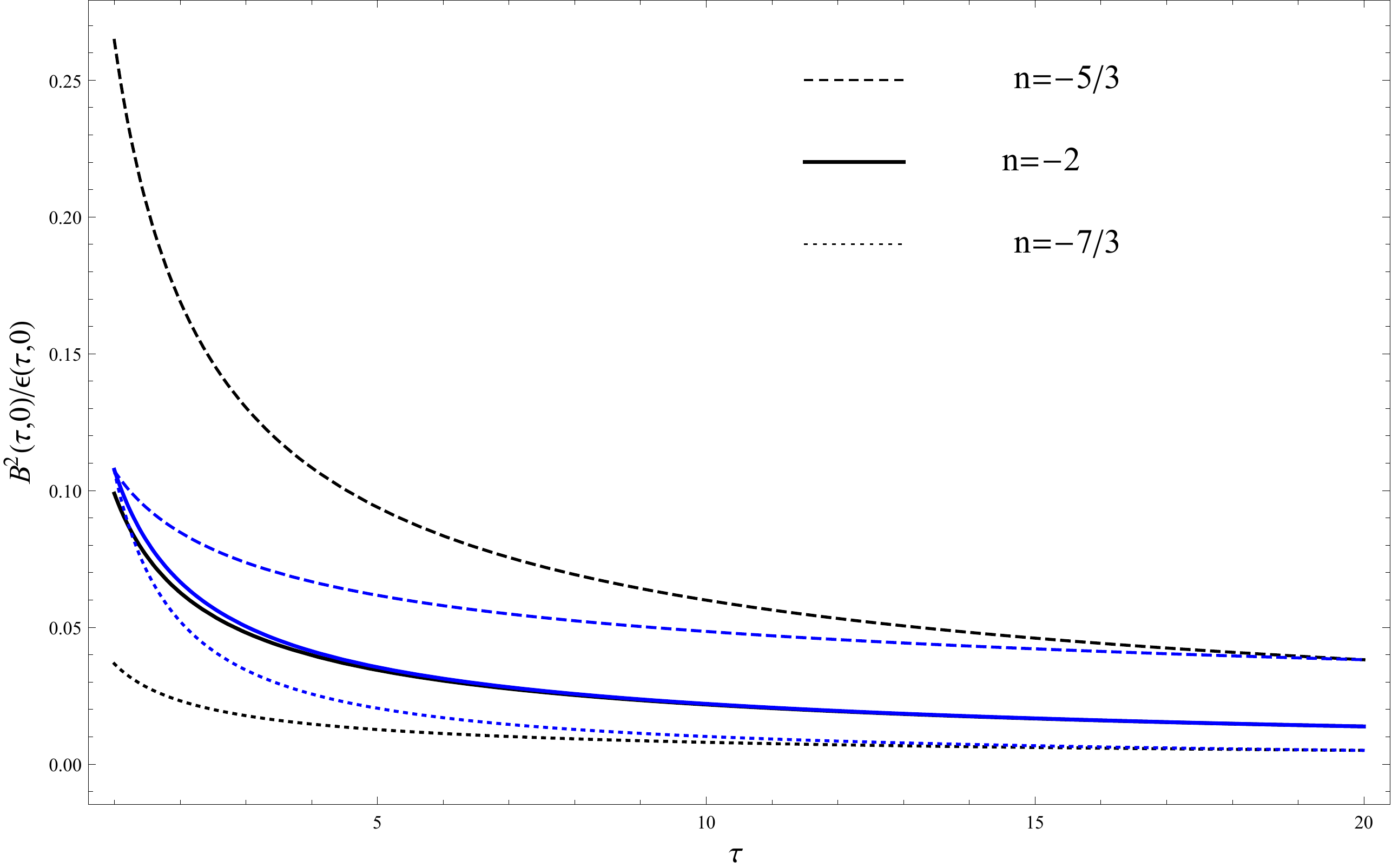}
\caption{$b(\tau, x)^2/\epsilon(\tau, x)$ versus $\tau$ plotted at $x=0$ with different values of $n$.
The black curves correspond to the analytical solutions of Ref.~\cite{pu}.}
\label{selfsim_plot0}}
\end{figure}

Fig.12 shows the ratio $b(\tau,0)^2/\epsilon(\tau,0)$ as a function of $\tau$ at $x=0$, for different
values of $n$: it is seen that all the analytical solutions of Ref.~\cite{pu} (blue lines) converge to
the value 0.1 at $\tau=1$ for any value of $n$, as expected. Our results, instead, reach the same limit only
for the value $n=2$, while for $n=-7/3$ ($n=-5/3$) the ratio is typically smaller (larger). This shows the effect
of the coupling between Maxwell's and conservation equations.
Moreover, for $n<-1$, the ratio decreases with increasing time: this implies that the energy density of the
magnetic field decays much faster than the fluid energy density in relativistic heavy ion collisions.

Finally, in order to show the influence of the RMHD  equations on the
modification of velocity, magnetic field and energy density we consider the specific case $n=-4/5$, which
corresponds to a weaker dependence on time of the magnetic field, Eq.~(\ref{magn}).
This is illustrated in Figs. 12 and 13, where again we compare our
solutions with the  approximated analytical ones of Ref.~\cite{pu}.  The latter implies that $v(\tau, x)=0$
for $n=-1$ , hence one may expect a change in the
direction of the transverse velocity.

Fig. 13 (a) and (b) show the transverse velocity results from the numerical solutions of the present work
and from the analytical solutions, respectively, at different times, for $n=-4/5$ .
From Fig.~13 (b) one finds that the magnitude of the transverse
velocity decreases with respect to the proper time $\tau_0=1$~fm, as expected, and the velocity profile has a similar
shape compared with the case $n<-1$, but the direction becomes negative. In addition (notice the small numbers
in the vertical scale) it is nearly zero, since in the approximate analytical solution the fluid velocity is
only modified by the spatial gradient of the external  magnetic field.
On the contrary, Fig.~13 (a) shows that the direction of fluid velocity is positive and decreases with time, until
$\tau=10$~fm, where the sign changes; moreover its magnitude is much larger
than the analytical solution  at early times.
As a conclusion we can state that, in the analytical solution, the transverse
flow led by a Gaussian magnetic field points outward for $n<-1$ and inward for $n>-1$, while the results of the
present work, where the RMHD equations  are solved numerically, for both cases ($n<-1$ and $n>-1$) the
transverse flow points outward in early time, though they have opposite direction at late time.

Fig.~14 (a) shows the behavior of  $b(\tau, 0)^2/\epsilon(\tau, 0)$  versus $x$, for different proper times:
it decreases with time from the value 2.5 at proper time $\tau_0=1$ fm  to the value 0.6 at the
late time $\tau_1=10$ fm, similarly to the case $ n<-1$ (see Fig.~12). We also plot the analytical
solution for $b(\tau, 0)^2/\epsilon(\tau, 0)$  in Fig.~14 (b): in this case the considered ratio increases
with time from the value $\tau_0=0.1$ at proper time $\tau_0=1$~fm to the value 0.6 at the late time $\tau_1=10$~fm.
This again stresses the importance of  the coupling between Maxwell's and conservation equations.

\begin{figure}[ht]
\center{\includegraphics[width=1\textwidth]{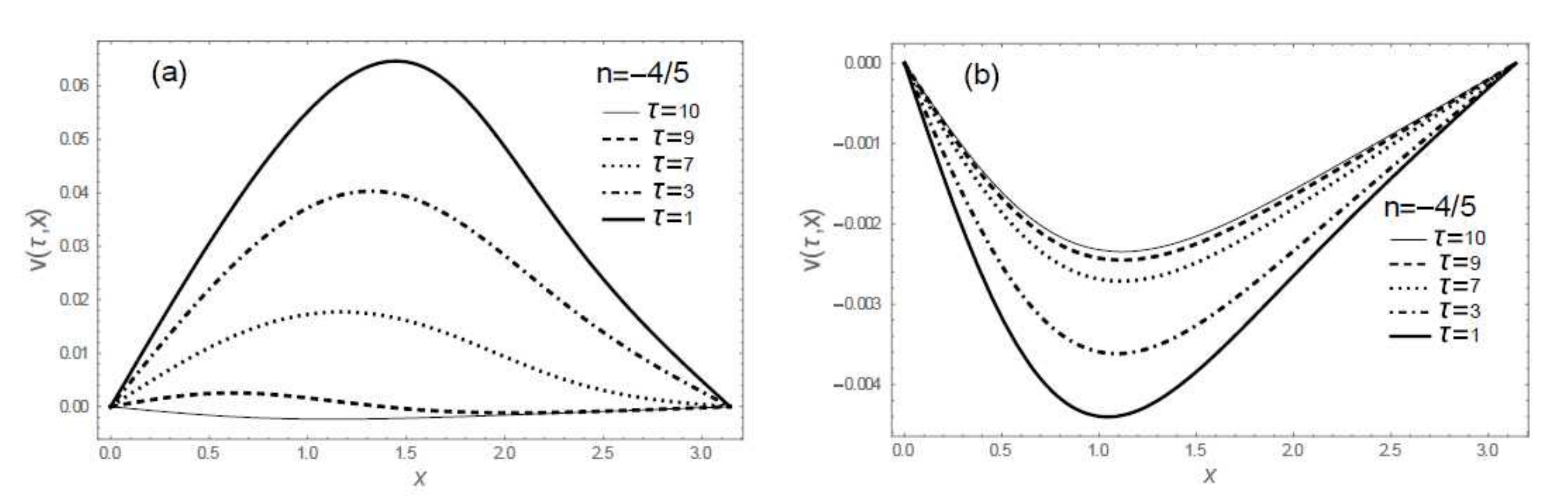}
\caption{Transverse velocity $v(\tau, x)$ versus $x$ plotted at different times for $n=-4/5$,
(a) numerical solutions
(present work), (b) analytical solutions of Ref.~\cite{pu}.}
\label{selfsim_plot1}}
\end{figure}
\begin{figure}[ht]
\center{\includegraphics[width=1\textwidth]{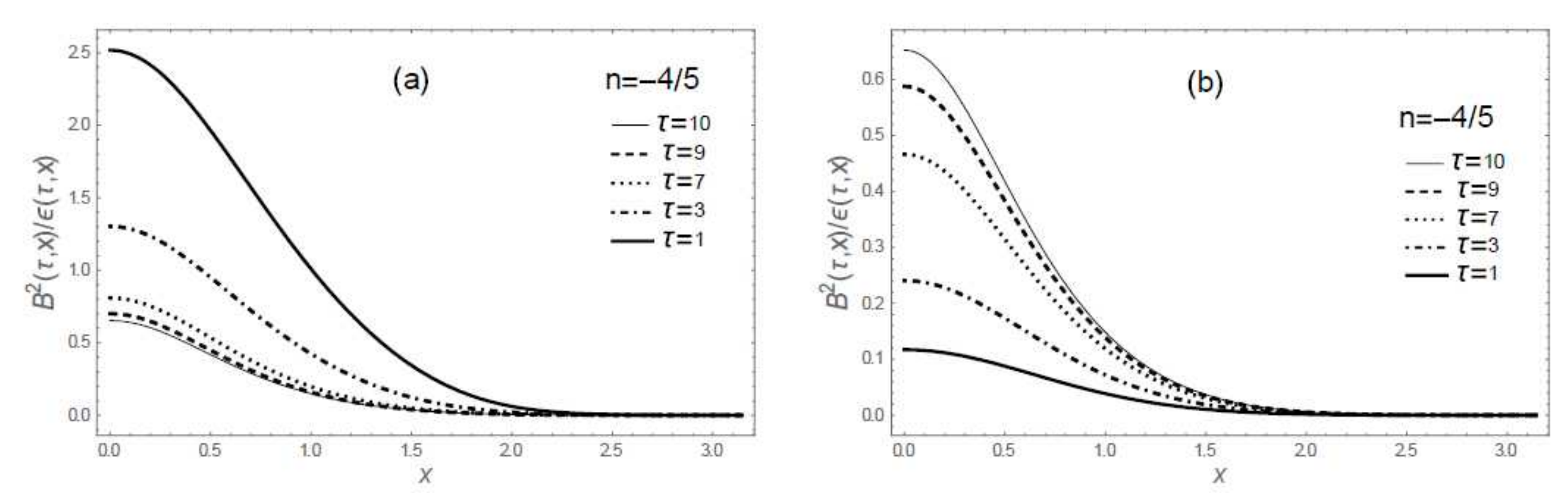}
\caption{$b(\tau, x)^2/\epsilon(\tau, x)$ versus $x$ plotted at different times for $n=-4/5$,
(a) numerical solutions (present work),
(b) analytical solutions .}
\label{selfsim_plot2}}
\end{figure}

 \section{Conclusion and outlook}

 In this work, we present a new numerical method for the solution of coupled relativistic hydrodynamic
equations and Maxwell’s equations, i.e., relativistic magnetohydrodynamics (RMHD), which has
recently become of  growing interest for the study of relativistic heavy ion collisions.
By solving  the coupled  conservation  and Maxwell’s
equations, we obtain numerical results for the fluid velocity, the energy density and the magnetic field.

We work in the 1+1 dimensional RMHD model where the transverse magnetic  field, fluid velocity, and energy
density are considered as a function of one spatial ($x$) and one temporal ($\tau$) variable;
besides, the magnetic field points along the orthogonal $y$ direction.
In our setup, the medium is boost-invariant along the $z$ direction.
 It turns out that the transverse velocity is rather small at all times and for different parameterizations
of the initial magnetic field: hence we treated the transverse flow in the non-relativistic approximation.

The core of our method is twofold: i) the adoption of a discretized spatial variable, in terms of which
the derivatives are expressed with the method of the second order finite difference formula, ii) the
adoption of precise and realistic boundary conditions for the numerical
solution of the resulting system of ordinary differential equations in the time variable.

Although it is known that during relativistic heavy ion collisions intense magnetic fields are developed,
their knowledge in the initial times of the collision, where the QGP has been formed, is very poor
and could not be used as the desired boundary conditions: hence we introduce initial conditions at late time
and solve numerically the coupled equations inversely in time.

We found it appropriate to assume for the late time quantities (fluid velocity, energy density and magnetic
field) the approximate analytical solutions found by the authors of Ref.~\cite{pu}: in contrast with our approach,
these solutions where found by neglecting the coupling between the dynamical evolution of the magnetic field and
the one of the energy density and velocity of the fluid. A weak external, uncoupled magnetic field is adopted, with
Gaussian distribution dependence in space and power-law decay dependence in time.
For our method these analytical solutions are appropriate for the late time boundary conditions, since
in the final stage of the plasma evolution the magnetic field is indeed quite small and its coupling can be
safely neglected.

After presenting our results for the fluid velocity, the energy density and the magnetic field as functions of both
space and time, we have validated our numerical calculation by making a comparison with the approximate analytical
solutions of Ref.~\cite{pu}. As expected the two approaches give similar results at late times as well as
for specific choices of the time-evolution of their magnetic field (which is governed by the parameter $n$),
but in other conditions the coupling between Maxwell and conservation (RMHD) equations, here taken into account,
appears to be quite relevant.

In particular we notice that the transverse velocities  have  the same direction in early and late time for the
case $n<-1$ (faster decay of the magnetic field), while for the case $n>-1$ the transverse velocities
appear to change sign in late time. Hence, the transverse flow propagates on the same direction,  for any value of $n$
only in the early stages of the collision. It should be noticed, however, that a weak decay of the magnetic field
is probably less realistic than a strong one.

According to the estimated conditions of heavy ion collision experiment at RHIC,
one find $b^2/\epsilon=0.17- 0.68$ at $\tau=0.6 $ fm. As a result, for the validity of the weak-field
expansion, in   Ref.~\cite{pu} $b^2/\epsilon=0.1$ was chosen at the proper time $\tau_0=1$.  We find that
only for the case $n=-2$, $b^2/\epsilon$ converges to this value at the proper time $\tau_0=1$. For the
case$ n<-2$, $b^2/\epsilon$ is smaller than 0.1 and decreases when  $n$ becomes smaller. For the case
$n>-2$, $b^2/\epsilon$ is bigger than 0.1 and increases when  $n$ becomes larger. For both cases
$n<-1$ and $n>-1$, $b(\tau, x)^2/\epsilon(\tau, x)$ decreases with time at fixed $x$.
This is a preliminary result of our approach, which is potentially interesting and deserves further investigation
also from experimental point of view.

As a final remark and outlook for future developments we wish to consider the possibility of extending the present
calculation to the case {\bf of} 2+1 dimensions, taking into account both transverse dimensions. This will also allow to
investigate differences in the azimuthal distribution of the velocities and hence the so-called elliptic flow.
It is well known that this is one of the crucial characteristics of the deconfined plasma.
At present there is an interesting debate about the influence of the magnetic field on the $v_2$ coefficient of
the elliptic flow (see for example ref.~\cite{Das,Roy}) and the role of the magnetic field in this connection still
deserves additional efforts.



\begin{thebibliography}{99}

\bibitem{Romatschke:2007mq}
  P.~Romatschke and U.~Romatschke,
  ``Viscosity Information from Relativistic Nuclear Collisions: How Perfect is the Fluid Observed at RHIC?,''
  Phys.\ Rev.\ Lett.\  {\bf 99} (2007) 172301.
\bibitem{Song:2007ux}
  H.~Song and U.~W.~Heinz,
  ``Causal viscous hydrodynamics in 2+1 dimensions for relativistic heavy-ion collisions,''
  Phys.\ Rev.\ C {\bf 77} (2008) 064901.
\bibitem{Bozek:2011ua}
  P.~Bozek,
  ``Flow and interferometry in 3+1 dimensional viscous hydrodynamics,''
  Phys.\ Rev.\ C {\bf 85} (2012) 034901.
\bibitem{Gale:2012rq}
  C.~Gale, S.~Jeon, B.~Schenke, P.~Tribedy and R.~Venugopalan,
  ``Event-by-event anisotropic flow in heavy-ion collisions from combined Yang-Mills and viscous 
fluid dynamics,''
  Phys.\ Rev.\ Lett.\  {\bf 110} (2013) no.1,  012302.
\bibitem{Andrade:2006yh}
  R.~Andrade, F.~Grassi, Y.~Hama, T.~Kodama and O.~Socolowski, Jr.,
  ``On the necessity to include event-by-event fluctuations in experimental evaluation of 
elliptical flow,''
  Phys.\ Rev.\ Lett.\  {\bf 97} (2006) 202302.
\bibitem{DelZanna:2013eua}
  L.~Del Zanna {\it et al.},
  ``Relativistic viscous hydrodynamics for heavy-ion collisions with ECHO-QGP,''
  Eur.\ Phys.\ J.\ C {\bf 73} (2013) 2524.

\bibitem{Kharzeev:2007jp} D.~E.~Kharzeev, L.~D.~McLerran and H.~J.~Warringa,
  ``The Effects of topological charge change in heavy ion collisions: 'Event by event P and CP violation',''
  Nucl.\ Phys.\ A {\bf 803}, 227 (2008).
\bibitem{Fukushima:2008xe}
  K.~Fukushima, D.~E.~Kharzeev and H.~J.~Warringa,
  ``The Chiral Magnetic Effect,''
  Phys.\ Rev.\ D {\bf 78}, 074033 (2008).
\bibitem{Voloshin:2004vk}
  S.~A.~Voloshin,
  ``Parity violation in hot QCD: How to detect it,''
  Phys.\ Rev.\ C {\bf 70} (2004) 057901
\bibitem{Abelev:2009ad}
  B.~I.~Abelev {\it et al.} [STAR Collaboration],
  ``Observation of charge-dependent azimuthal correlations and possible local strong parity violation 
in heavy ion collisions,''
  Phys.\ Rev.\ C {\bf 81} (2010) 054908.
\bibitem{Abelev:2009ac}
  B.~I.~Abelev {\it et al.} [STAR Collaboration],
  ``Azimuthal Charged-Particle Correlations and Possible Local Strong Parity Violation,''
  Phys.\ Rev.\ Lett.\  {\bf 103} (2009) 251601.
\bibitem{Adam:2015vje}
  J.~Adam {\it et al.} [ALICE Collaboration],
  ``Charge-dependent flow and the search for the chiral magnetic wave in Pb-Pb collisions 
at $\sqrt{s_{\rm NN}} =$ 2.76 TeV,''
  Phys.\ Rev.\ C {\bf 93} (2016) no.4,  044903
\bibitem{Li:2014bha}
  Q.~Li {\it et al.},
  ``Observation of the chiral magnetic effect in ZrTe5,''
  doi:10.1038/nphys3648
  arXiv:1412.6543 [cond-mat.str-el].
\bibitem{Xiong:2015nna}
  J.~Xiong, S.~K.~Kushwaha, T.~Liang, J.~W.~Krizan, W.~Wang, R.~J.~Cava and N.~P.~Ong,
  ``Signature of the chiral anomaly in a Dirac semimetal: a current plume steered by a magnetic field,''
  arXiv:1503.08179 [cond-mat.str-el].
\bibitem{Huang:2015eia}
  X.~Huang {\it et al.},
  ``Observation of the Chiral-Anomaly-Induced Negative Magnetoresistance in 3D Weyl Semimetal TaAs,''
  Phys.\ Rev.\ X {\bf 5} (2015) no.3,  031023.
\bibitem{Shekhar:2015rqa}
  C.~Shekhar {\it et al.},
  ``Large and unsaturated negative magnetoresistance induced by the chiral anomaly in the Weyl 
semimetal TaP,''
  arXiv:1506.06577 [cond-mat.mtrl-sci].
\bibitem{Kharzeev:2015znc}
  D.~E.~Kharzeev, J.~Liao, S.~A.~Voloshin and G.~Wang,
  ``Chiral magnetic and vortical effects in high-energy nuclear collisions—A status report,''
  Prog.\ Part.\ Nucl.\ Phys.\  {\bf 88}, 1 (2016).

\bibitem{a2} Claudio Bonati, Massimo D’Elia, Marco Mariti, Michele Mesiti, and Francesco Negro,
" Anisotropy of the quark-antiquark potential in a magnetic field ", Phys. Rev D 89, 114502 (2014).

\bibitem{a3} G. S. Balia, F. Bruckmann, G. Endrodi, Z.Fodor, S.D.Katz, A.Schäfer, 
" Effects of magnetic fields on the quark–gluon plasma ", Nucl. Phys. A 931 (2014) 752–757.

\bibitem{a4} Andersen - Phys. Rev. D 86 (2012), Luschevskaya and Larina - JETP Letters 98 (2014)

\bibitem{a5} Bali, Bruckmann, Endrodi et al. - Journal of High Energy Physics 08 177 (2014).

\bibitem{a6} R.Belmontfor the ALICE Collaboration, " Charge-dependent anisotropic flow studies and the 
search for the Chiral Magnetic Wave in ALICE ", Nucl. Phys. A 931 (2014) 981–985.

\bibitem{a7} Wei-Tian Deng, Xu-Guang Huang, " Electric fields and chiral magnetic effect in 
Cu+Au collisions ", Phys. Lett. B 742 (2015) 296–302.

\bibitem{a9} K. Tuchin, " Electromagnetic field and the chiral magnetic effect in the quark-gluon 
plasma ", arXiv:1411.1363v1 [hep-ph].

\bibitem{a11} R. Belmontfor the ALICE Collaboration, " Charge-dependent anisotropic flow studies and 
the search for the Chiral Magnetic Wave in ALICE ", Nucl. Phys. A 931 (2014) 981–985.

\bibitem{Tuchin:2013apa}
  K.~Tuchin,
  ``Time and space dependence of the electromagnetic field in relativistic heavy-ion collisions,''
  Phys.\ Rev.\ C {\bf 88} (2013) no.2,  024911.

\bibitem{Tuchin:2013ie}
  K.~Tuchin,
  ``Particle production in strong electromagnetic fields in relativistic heavy-ion collisions,''
  Adv.\ High Energy Phys.\  {\bf 2013} (2013) 490495.

\bibitem{a17} K. Tuchin, " Electromagnetic fields in high energy heavy-ion collisions ", 
Int. J.  Mod. Phys. E Vol. 23, No. 1 (2014) 1430001.

\bibitem{a20} B. G. Zakharov, " Electromagnetic response of quark gluon plasma in heavy ion collisions ", 
Phys. Lett. B 737 (2014) 262-266.
\bibitem{McLerran:2013hla}
  L.~McLerran and V.~Skokov,
  ``Comments About the Electromagnetic Field in Heavy-Ion Collisions,''
  Nucl.\ Phys.\ A {\bf 929} (2014) 184.

\bibitem{Deng:2012pc}
  W.~T.~Deng and X.~G.~Huang,
  ``Event-by-event generation of electromagnetic fields in heavy-ion collisions,''
  Phys.\ Rev.\ C {\bf 85} (2012) 044907.

\bibitem{chinees} H. Li, X-L. Sheng, Q. Wang, " Electromagnetic fields with electric and 
chiral magnetic conductivities in heavy ion collisions ", 
Phys. Rev.  C { \bf 94}, 044903 (2016)

\bibitem{bj83} J. D. Bjorken, Phys. Rev. D 27, 140 (1983).

\bibitem{Gursoy:2014aka}
  U.~Gursoy, D.~Kharzeev and K.~Rajagopal,
  ``Magnetohydrodynamics, charged currents and directed flow in heavy ion collisions,''
  Phys.\ Rev.\ C {\bf 89} (2014) no.5,  054905

\bibitem{Bzdak:2011yy}
  A.~Bzdak and V.~Skokov,
  ``Event-by-event fluctuations of magnetic and electric fields in heavy ion collisions,''
  Phys.\ Lett.\ B {\bf 710} (2012) 171.

\bibitem{a14} V. V. Skokov, A. Yu. Illarionov and V. D. Toneev ," Estimate of the magnetic field strength 
in heavy-ion collision ", 
Int. J.  Mod. Phys. A Vol. 24, No. 31 (2009) 5925–5932.

\bibitem{a15} Yang Zhong, Chun-Bin Yang, Xu Cai, and Sheng-Qin Feng, 
" A Systematic Study of Magnetic Field in Relativistic Heavy-Ion Collisions in the RHIC and LHC 
Energy Regions ", 
Advances in High Energy Physics Volume 2014, Article ID 193039.

\bibitem{a23} V. Voronyuk, V. D. Toneev, W. Cassing, E. L. Bratkovskaya, V. P. Konchakovski, S. A. Voloshin,"
 Electromagnetic field evolution in relativistic heavy-ion collisions ", 
Phys. Rev C 83, 054911 (2011).

\bibitem{a27} Victor Roy, Shi Pu, " Event-by-event distribution of magnetic field energy over 
initial fluid energy density
in $\sqrt{S_{NN}}= 200$ GeV Au-Au collisions ", Phys. Rev C 92, 064902, (2015).



\bibitem{a29}J. Goedbloed, R. Keppens, S. Poedts"  Advanced Magnetohydrodynamics with Applications 
to Laboratory and Astrophysical Plasmas ", Cambridge University Press, 2010.

\bibitem{an89} A. M. Anile, " Relativistic fluids and magneto-fluids ", Cambdridge University Press (1989).

\bibitem{roy15} V. Roy, S. Pu, L. Rezzolla, D. Rischke, " Analytic Bjorken flow in one-dimensional 
relativistic magnetohydrodynamics ", Physics Letters B, Vol. 750, (2015).

\bibitem{pu16} S. Pu, V. Roy, L. Rezzolla and D. H. Rischke, " Bjorken flow in one-dimensional 
relativistic magnetohydrodynamics with magnetization ", Phys. Rev. D 93, 074022 (2016).

\bibitem{pang16} L. G. Pang, G. Endr\"{o}di and H. Petersen, " Magnetic field-induced squeezing 
effect at RHIC and at the LHC ", Phys. Rev. C 93, 044919 (2016).

\bibitem{pu} Shi Pu, and Di-Lun Yang, '' Transverse flow induced by inhomogeneous magnetic fields
in the Bjorken expansion '', Phys. Rev D 93, 054042 (2016).

 \bibitem{Inghirami etal} G. Inghirami, L. Del Zanna, A. Beraudo, M. Haddadi Moghaddam, 
F. Becattini, M. Bleicher, " Numerical magneto-hydrodynamics for relativistic nuclear collisions ",  
Eur. Phys. J. C (2016) 76:659.

\bibitem{Das} A.~Das, S.S.~Dave, P.S.~Saumia and A. M. Srivastava, ``Effects of magnetic field on the 
plasma evolution in relativistic heavy-ion collisions'', arXiv:1703.08162 [hep-ph]



\end{thebibliography}
\end{document}